\documentclass[conference]{IEEEtran}

\IEEEoverridecommandlockouts   

\usepackage[mode=buildmissing]{standalone} % 
\usepackage{amsmath}
\usepackage{bm}
\usepackage[nolist]{acronym}
\usepackage{amssymb}
\usepackage{comment}
\usepackage{graphicx}
\usepackage{color,colortbl}
\usepackage{xcolor}
\usepackage{cleveref}
\usepackage{verbatim}
\usepackage{enumerate}
\usepackage[shortlabels]{enumitem}
\usepackage{cuted, nccmath}
\usepackage{dblfloatfix}
\usepackage{float} 
\usepackage{lipsum}
\usepackage{balance}
\usepackage{cite}
\usepackage{url}
\usepackage{tabularx}
\usepackage{boldline}
\usepackage{mathtools} 
\usepackage{stmaryrd}
\usepackage{graphicx}
\usepackage{pgfplots}
\usepackage{booktabs}
\usepackage[font={footnotesize}]{caption}
\usepackage{subfig} 
\usepackage{tikz}
\usepackage{xparse} 
\usepackage{algorithm}
\usepackage{algpseudocode}
\usepackage[all=normal,paragraphs=tight,floats=normal,mathspacing=normal,wordspacing=tight,charwidths=tight,mathdisplays=normal,leading=normal]{savetrees}

\usetikzlibrary{arrows.meta}
\tikzset{every picture/.style={line width=0.6pt}}
\usetikzlibrary{arrows.meta,
                calc, 
                backgrounds,
                chains,
                fit,
                quotes,
                shapes.geometric,
                positioning,
                intersections,
                spath3,
                spy}

\algdef{SE}[SUBALG]{Indent}{EndIndent}{}{\algorithmicend\ }%
\algtext*{Indent}
\algtext*{EndIndent}

\DeclareDocumentCommand{\complexset}{o o}{%
    \mathbb{C}\IfValueT{#1}{\IfValueTF{#2}{^{#1\times#2}}{^{#1}}}
}
\DeclareDocumentCommand{\realset}{o o}{%
    \mathbb{R}\IfValueT{#1}{\IfValueTF{#2}{^{#1\times#2}}{^{#1}}}
}
\newcommand{\hermit}{\mathsf{H}}

\newcommand{\dR}{d_{\mathrm{R}}}
\newcommand{\Tsym}{T_\mathrm{sym}}
\newcommand{\Tcp}{T_\mathrm{cp}}
\newcommand{\Deltaf}{\Delta_\mathrm{f}}
\newcommand{\sums}{\sum_{s=0}^{S-1}}
\newcommand{\rect}{\mathrm{rect}}

\newcommand{\arx}{\bm{a}}

\newcommand{\bmb}{\bm{b}}

\newcommand{\bmY}{\bm{Y}}
\newcommand{\bmZ}{\bm{Z}}
\newcommand{\bmx}{\bm{x}}
\newcommand{\bmN}{\bm{N}}
\renewcommand{\vec}{\mathrm{vec}}
\newcommand{\cnormal}{\mathcal{CN}}
\newcommand{\bmI}{\bm{I}}
\newcommand{\E}{\mathbb{E}}

\newcommand{\zn}{z_n}

\DeclareDocumentCommand{\bmkappa}{o}{%
    \bm{\kappa}\IfValueT{#1}{^{(#1)}}
}

\newcommand{\bmYtilde}{\widetilde{\bm{Y}}}

\newcommand{\bmM}{\bm{M}}
\newcommand{\Hnull}{\mathcal{H}_0}
\newcommand{\Halt}{\mathcal{H}_1}
\newcommand{\Lcal}{\mathcal{L}}
\newcommand{\Llog}{\mathcal{L}_{\mathrm{log}}}
\newcommand{\sigmagamma}{\sigma_{\gamma}}

\newcommand{\norm}[1]{\lVert#1\rVert}
\newcommand{\U}{\mathcal{U}}
\newcommand{\etatilde}{\tilde{\eta}}
\newcommand{\etabar}{\bar{\eta}}
\newcommand{\gammahat}{\hat{\gamma}}
\newcommand{\Oset}{\mathcal{O}}
\newcommand{\Tset}{\mathcal{T}}

\newcommand{\degree}{^\circ}
\newcommand{\thetamin}{\theta_{\min}}
\newcommand{\thetamax}{\theta_{\max}}
\newcommand{\taumin}{\tau_{\min}}
\newcommand{\taumax}{\tau_{\max}}
\newcommand{\thetamean}{\theta_{\mathrm{mean}}}
\newcommand{\Deltatheta}{\Delta_{\theta}}
\newcommand{\Rmin}{R_{\min}}
\newcommand{\Rmax}{R_{\max}}

\newcommand{\SNR}{\mathrm{SNR}}

\newcommand{\thetahat}{\hat{\theta}}
\newcommand{\tauhat}{\hat{\tau}}
\newcommand{\bmW}{\bm{W}}

\newcommand{\Ntheta}{N_{\theta}}
\newcommand{\Ntau}{N_{\tau}}
\newcommand{\bmPhitheta}{\bm{\Phi}_{\theta}}
\newcommand{\bmPhitau}{\bm{\Phi}_{\tau}}

\newcommand{\bmkappahat}{\hat{\bmkappa}}

\newcommand{\Jcal}{\mathcal{J}}

\newcommand{\hdet}{ \underset{\mathcal{H}_0}{\overset{\mathcal{H}_1}{\gtrless}} }

\begin{acronym}[ACRONYM]
\acro{6G}{6th generation wireless systems}
\acro{AWGN}{additive white Gaussian noise}
\acro{AOA}{angle-of-arrival}

\acro{CP}{cyclic prefix}
\acro{CSI}{channel state information}

\acro{DL}{deep learning}
\acro{GPI}{gain-phase impairment}
\acro{ISAC}{integrated sensing and communication}

\acro{LOS}{line-of-sight}

\acro{MAPRT}{maximum a-posteriori ratio test}
\acro{MB-ML}{model-based machine learning}
\acro{MIMO}{multiple-input multiple-output}
\acro{MISO}{multiple-input single-output}

\acro{OFDM}{orthogonal frequency-division multiplexing}

\acro{QPSK}{quadrature phase-shift keying}

\acro{TX}{transmit}

\acro{RX}{receive}

\acro{SIMO}{single-input multiple-output}
\acro{SL}{supervised learning}
\acro{SNR}{signal-to-noise ratio}

\acro{UE}{user equipment}
\acro{UL}{unsupervised learning}
\acro{ULA}{uniform linear array}
\end{acronym}

\begin{document}
\bstctlcite{IEEEexample:BSTcontrol}

\title{Unsupervised Learning for Gain-Phase Impairment\\ Calibration in ISAC Systems \\
\thanks{This work was supported, in part, by a grant from the Chalmers AI Research Center Consortium (CHAIR), by the National Academic Infrastructure for Supercomputing in Sweden (NAISS), the Swedish Foundation for Strategic Research (SSF) (grant FUS21-0004, SAICOM), Hexa-X-II, part of the European Union’s Horizon Europe research and innovation programme under Grant Agreement No 101095759, and Swedish Research Council (VR grant 2022-03007). The work of C.~Häger was also supported by the Swedish Research Council under grant no. 2020-04718. The work of L.~Le Magoarou is supported by the French national research agency (grant ANR-23-CE25-0013)}
}
\author{Jos\'{e} Miguel Mateos-Ramos\IEEEauthorrefmark{1}, Christian H\"{a}ger\IEEEauthorrefmark{1}, Musa Furkan Keskin\IEEEauthorrefmark{1}, \\ Luc Le Magoarou\IEEEauthorrefmark{2}, Henk Wymeersch\IEEEauthorrefmark{1}\\
\IEEEauthorrefmark{1}Department of Electrical Engineering, Chalmers University of Technology, Sweden \\
\IEEEauthorrefmark{2}Univ Rennes, INSA Rennes, CNRS, IETR-UMR 6164, Rennes, France}

\maketitle

\IEEEpeerreviewmaketitle

\begin{abstract}
\Acp{GPI} affect both communication and sensing in 6G integrated sensing and communication (ISAC). 
We study the effect of \acp{GPI} in a single-input, multiple-output orthogonal frequency-division multiplexing ISAC system and develop a model-based unsupervised learning approach to simultaneously (i) estimate the gain-phase errors and (ii)  localize sensing targets. 
The proposed method is based on the optimal maximum a-posteriori ratio test for a single target. Results show that the proposed approach can effectively estimate the gain-phase errors and yield similar position estimation performance as the case when the impairments are fully known.
\end{abstract}

\acresetall

\section{Introduction}
\Ac{ISAC} is considered a key enabler of the \ac{6G} \cite{cui2023integrated}, combining sensing and communication functions in a single device, thereby providing sensing capabilities to communication systems, while also improving wireless channel usage efficiency and system performance \cite{liu2022integrated}.
Signal processing in ISAC has been largely driven by model-based algorithms, which offer performance guarantees, explainability, and predictable computational complexity \cite{liyanaarachchi2021joint, dokhanchi2021adaptive, OFDM_DFRC_TSP_2021, chen2021joint, johnston2022mimo}. 
However, the higher carrier frequencies expected in 6G and the integration of sensing in communication networks increase the likelihood of hardware impairments such as antenna distortions, phase noise, and sampling jitter \cite{tan2021integrated, alsabah20216g}. These hardware impairments cause a model mismatch in the model-based algorithms and thus degrade their performance. 

\Ac{DL}  has been successfully applied to mitigate hardware impairments in \ac{ISAC}~\cite{liu2022learning, wu2023sensing, liu2023deep, charlotte2024loss}, but it suffers from lack of interpretability. In contrast, \ac{MB-ML}  provides interpretable solutions,  by  parameterizing standard model-based algorithms, enhancing their adaptability to mismatched models while offering performance guarantees~\cite{Shlezinger2023model}. 
MB-ML has been applied in communications \cite{Hengtao18, Xiuhong21, yassine2022mpnet, chatelier2023efficient}, sensing \cite{xiao2020deepfpc, wu2022doa, shmuel2023subspacenet}, and ISAC scenarios \cite{mateos2023model}. 
Hardware impairment mitigation solutions (e.g., \cite{wu2022doa, shmuel2023subspacenet, mateos2023model}) rely on \ac{SL}, which involves the difficult or time-consuming process of acquiring the ground-truth position of the objects in the environment. 
\Ac{UL} avoids labeled data and has been applied for ISAC inter-antenna spacing impairment mitigation in \cite{mateos2024semi}, though still requiring a small labeled dataset to fully compensate for the impairments. 
\begin{figure}[tb]
    \centering
    \includegraphics[width=0.45\textwidth]{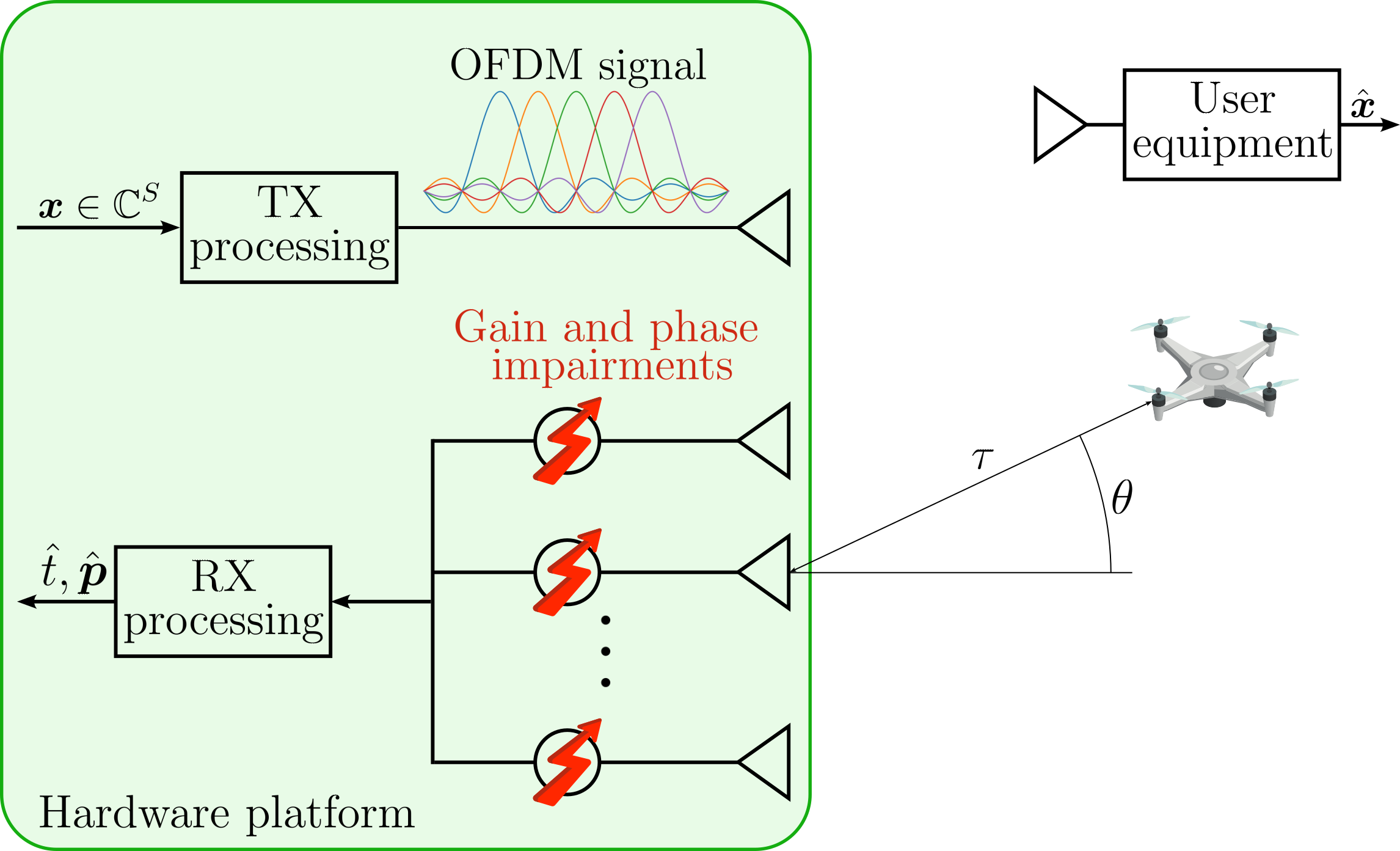}
    \caption{A monostatic SIMO radar sends OFDM signals to perform single-target detection and position estimation. The OFDM signals can be received by a user equipment to estimate the transmitted messages and reflected back from a target in the environment. The receive antenna array is affected by GPIs.}
    \label{fig:model}
\end{figure}

In this paper, we develop an UL approach to jointly compensate for antenna \acp{GPI} and estimate target locations, under the MB-ML framework. As a proof-of-concept, we focus on a simple monostatic \ac{SIMO} radar performing single-target detection and position estimation based on \ac{OFDM} signals (see   Fig.~\ref{fig:model}). 
\ac{GPI} mitigation is an important problem, with model-based  \cite{ng2009practical, liu2023new, li2019blind}, \ac{DL}  \cite{jie2022gain, zhou2023transfer}, and MB-ML solutions \cite{yassine2022mpnet, chatelier2023efficient}. However, \cite{ng2009practical, liu2023new} required at least a transmitter at a known angle to calibrate the antenna array and \cite{li2019blind} assumed a known model of the \ac{CSI}, which does not apply to our case as the \ac{CSI} contains the target position to be estimated. 
In \cite{jie2022gain}, only angle estimation was performed and \cite{zhou2023transfer} required a transmitter at a known position for calibration. Moreover, \cite{jie2022gain, zhou2023transfer} require labeled data to train.
Finally, although \cite{yassine2022mpnet, chatelier2023efficient} proposed MB-ML UL approaches to compensate for GPI, they considered a pure communication system and CSI estimation without localization of the user.

\section{System Model}
We consider a monostatic \ac{SIMO}-\ac{OFDM} ISAC transceiver equipped with a single-antenna \ac{TX} and a \ac{RX} \ac{ULA} of $N$ elements. The element spacing of the RX ULA is $\dR$. The \ac{OFDM} signal has a symbol duration of $\Tsym = \Tcp + T$, where $\Tcp$ is  the \ac{CP} and and $T$ is the elementary symbol duration. 
The complex baseband OFDM transmit signal with $S$ subcarriers and a subcarrier spacing  $\Deltaf=1/T$ is \cite{5G_NR_JRC_analysis_JSAC_2022,OFDM_radar_TVT_2020}
\begin{align} \label{eq:s_t}
    s(t) = \frac{1}{\sqrt{S}} \sums x_s e^{\jmath 2\pi s\Deltaf  t} \rect\bigg(\frac{t}{\Tsym}\bigg),
\end{align}
where $x_s$ is the complex transmitted symbol in the $s$-th subcarrier. 
Considering the presence of a stationary point-target in the far-field, the noise-free received baseband signal at the $n$-th RX element is \cite{MIMO_OFDM_ICI_JSTSP_2021}
\begin{align} \label{eq:ymn_t_2}
    \zn(t) = &{\gamma} [\arx(\theta)]_n s (t-\tau)
\end{align}
where $\gamma$ is the complex channel gain, $\tau$ is the  total round-trip delay of the target, $\theta$ is the \ac{AOA}, and $\arx(\theta)$ is the array steering vector with 
\begin{align}
    [\arx(\theta)]_n &= e^{ -\jmath 2 \pi  {n\dR\sin(\theta)}/{\lambda} }, n=0,\ldots,N-1
\end{align}
for carrier wavelength $\lambda$. 
Following the standard OFDM assumption, the CP is taken to be larger than the round-trip delay of the furthermost target, i.e., $\Tcp \geq \tau$.
Sampling $\zn(t)$ at $t = \Tcp + lT/S$ for $l=0,\ldots,S-1$ (i.e., after CP removal), we obtain the discrete-time signal 
\begin{align} \label{eq:ymn_t_sampled}
    \zn[l] = & \frac{\gamma}{\sqrt{S}} [\arx(\theta)]_n \sums x_s e^{\jmath 2\pi s \frac{l}{S}} e^{-\jmath 2\pi s \Deltaf \tau},
\end{align}
where the known phase shift $\exp(\jmath 2\pi s \Deltaf \Tcp)$ is absorbed into $x_s$. 
Taking the $S$-point DFT of $\zn[l]$ yields the frequency-domain baseband signal as
\begin{align} \label{eq:Y_mn_s_1}
    Z_{{n,s}} &= \mathcal{F}\big\{ \{\zn[l]\}_{l=0}^{S-1} \big\} = \gamma [\arx(\theta)]_n x_{s} [\bmb(\tau)]_{s},
\end{align}
with$[\bmb(\tau)]_s = \exp(-\jmath 2\pi s \Deltaf \tau)$. Aggregating over antenna elements and subcarriers, the signal in \eqref{eq:ymn_t_sampled} can be expressed as
\begin{align} \label{eq:Y_matrix}
    \bmZ = &\gamma \arx(\theta) (\bmb(\tau) \odot \bmx)^\top \in \complexset[N][S],
\end{align}
where $\bmx = [x_0 \cdots x_{S-1}]^\top$ is the transmit symbol vector and $\odot$ denotes the Hadamard product.

\subsubsection*{Observation without \ac{GPI}}
Adding noise at the receiver side and considering the random presence of a target in the environment yields the final model\footnote{The communication receiver is not affected by \ac{GPI} under the considered \ac{SIMO} model. For this reason and due to space limitations, the communication performance is not evaluated in this paper. }
\begin{align} \label{eq:Y_noisy}
    \bmY = &t\gamma \arx(\theta) (\bmb(\tau)\odot \bmx)^\top + \bmW,
\end{align}
where $t\in\{0,1\}$ denotes the absence or presence of a target and $\bmW$ represents \ac{AWGN} following $\vec(\bmW)\sim\cnormal(\bm{0}, N_0\bm{I})$, with $\vec(\cdot)$ the vectorization operation, $\bm{0}$ the all-zeros vector and $\bm{I}$ the identity matrix. The goal of the sensing receiver is to detect the presence of the target and estimate its position based on $\bmY$.

\subsubsection*{Observation with \ac{GPI}}
When the ULA elements are affected by \acp{GPI}, the actual steering vector of the ULA is $\arx(\theta;\bmkappa) = \bmkappa\odot \arx(\theta)$, where $\bmkappa \in \complexset[N]$ is a vector that contains the \acp{GPI} of all antenna elements. We consider that $\norm{\bmkappa}^2=N$ so that under impairments the transmitter energy is preserved, i.e., $\norm{\arx(\theta;\bmkappa)}^2 = \norm{\arx(\theta)}^2=N$. The model in \eqref{eq:Y_noisy} under \acp{GPI} becomes
\begin{align} \label{eq:Y_final}
    \bmY = &t\gamma (\arx(\theta;\bmkappa) (\bmb(\tau)\odot \bmx)^\top) + \bmW.
\end{align}
The goal of the receiver is now to operate  under unknown $\bmkappa$.

\section{Proposed Method} 
In the following, we detail the considered baseline to perform target detection and position estimation as well as the proposed unsupervised MB-ML approach to compensate for the \acp{GPI}. 
\subsection{Baseline}\label{sec:baseline}
We assume that the baseline operates under a fixed $\bmkappa$, which may not coincide with the true \acp{GPI}.
In order to detect the target presence and estimate its position, we resort to the \ac{MAPRT} detector \cite{MAP_Detector_TSP_2021}, which generalizes the generalized likelihood ratio test detector \cite{glrt_2001} to the case with random parameters and thus can take into account prior information on $\gamma$, $\theta$ and $\tau$. We assume that the complex channel gain follows a normal distribution as $\gamma\sim\cnormal(0,\sigmagamma^2)$ and the target angle and range are confined to an a priori known region, i.e., $\theta\sim\U[\thetamin, \thetamax]$, $\tau\sim\U[\taumin, \taumax]$. Moreover, we assume that $p(t=0)=p(t=1)=1/2$. For a fixed $\bmkappa$, the MAPRT then yields the following optimal test:
\begin{align} \label{eq:opt_test}
    \max_{\substack{\theta\in\Oset \\ \tau\in\Tset}} \big\{ |\arx^\hermit(\theta;\bmkappa) \bmY (\bmb(\tau) \odot \bmx)^\ast|^2 \big\} \gtrless \eta,
\end{align}
where $\Oset=[\thetamin, \thetamax]$, $\Tset=[\taumin, \taumax]$, $(\cdot)^\hermit$ denotes the conjugate transpose operation, $(\cdot)^*$ denotes the conjugate operation, $|\cdot|$ denotes the absolute value, and $\eta$ is a threshold that controls the probabilities of detection and false alarm. Details about the derivation of the MAPRT can be found in Appendix~\ref{app:maprt}. The angle and delay of the target are obtained as follows:
\begin{align} \label{eq:argmax_theta_tau}
    (\thetahat, \tauhat) = \arg\max_{\substack{\theta\in\Oset \\ \tau\in\Tset}} \big\{ |\arx^\hermit(\theta;\bmkappa) \bmY (\bmb(\tau) \odot \bmx)^\ast|^2 \big\}.
\end{align}
When the assumed $\bmkappa$ matches the actual GPIs, the baseline in \eqref{eq:opt_test}, \eqref{eq:argmax_theta_tau} is optimal and it represents a lower bound on the performance, as it will be shown in Sec.~\ref{sec:results}.

\subsection{Proposed \ac{UL} \ac{MB-ML} Method} \label{sec:proposed_method}
We base our approach on the baseline of Sec.~\ref{sec:baseline}. 
In particular, we compute the \textit{angle-delay} map as 
\begin{align} \label{eq:ad_map}
    \bmM(\bmkappahat) = |\bmPhitheta(\bmkappahat)^\hermit \bmY (\bmPhitau \odot \bmx \bm{1}^\top)^*|^2,
\end{align}
where $\bmkappahat$ is the estimate of the \acp{GPI}, $\bm{1}$ is the all-ones vector, and
\begin{align}
    \bmPhitheta(\bmkappahat) &= [\arx(\theta_1;\bmkappahat)\ \arx(\theta_2;\bmkappahat)\ \cdots\ \arx(\theta_{\Ntheta};\bmkappahat)] \label{eq:bmPhi_theta} \\
    \bmPhitau &= [\bmb(\tau_1)\ \bmb(\tau_2)\ \cdots\ \bmb(\tau_{\Ntau})]. \label{eq:bmPhi_tau}
\end{align}
We evaluate the angle-delay map on a uniformly sampled 2D grid, with $\Ntheta$ and $\Ntau$ the number of angle and delay points, respectively.
From the angle-delay map, we propose two different unsupervised loss functions to learn the \acp{GPI}.
\subsubsection{Maximize the maximum value of the angle-delay map}
\begin{align}
    \Jcal(\bmkappahat) &= \E_{t,\gamma,\theta,\tau, \bmx,\bmW}\big[ -\max_{i,j} [\bmM(\bmkappahat)]_{i,j}\big], \label{eq:loss_max}
\end{align}
where the expectation is taken with respect to random realizations of $t, \gamma, \theta, \tau, \bmx$, and $\bmW$ in \eqref{eq:Y_final}.
Unknown \acp{GPI} reduce the magnitude of the angle-delay map, since computation of the angle-delay map involves $|\arx^\hermit(\theta;\bmkappahat)\arx(\theta;\bmkappa)|^2$, which is only maximized if $\bmkappahat=\bmkappa$. Thus, we expect that by minimizing \eqref{eq:loss_max}, our proposed algorithm converges to the true impairments $\bmkappa$. Details about how the impairments affect the angle-delay map will be shown in Sec.~\ref{sec:results}. 

\subsubsection{Minimize the error of the received observation signal}
\begin{align}
    \Jcal(\bmkappahat) = \E_{t,\gamma,\theta,\tau, \bmx,\bmW}\big[\norm{\bmY - \bmYtilde(\bmkappahat)}_F\big], \label{eq:loss_norm}
\end{align}
where $\norm{\cdot}_F$ denotes the Frobenius norm and 
\begin{align} \label{eq:Y_reconstructed}
    \bmYtilde(\bmkappahat) = \gammahat \arx(\thetahat;\bmkappahat) (\bmb(\tauhat) \odot \bmx)^\top
\end{align}
is the reconstructed observation from the channel gain, angle, and target delay estimations. The expression for $\gammahat$ is derived in Appendix~\ref{app:maprt}.\footnote{The estimation of $\gammahat$ assumes knowledge of $N_0/\sigmagamma^2$, which is related to the \ac{SNR}. In this work we assume perfect knowledge of $N_0/\sigmagamma^2$, but we refer the reader to \cite{Pauluzzi2000comparison, wiesel2006snr} for SNR estimation methods.}
The motivation behind \eqref{eq:loss_norm} is that the observation $\bmY$ is affected by the true \acp{GPI} $\bmkappa$, while we reconstruct the observation $\bmYtilde$ in \eqref{eq:Y_reconstructed} using the estimated impairments $\bmkappahat$. Our hypothesis is that by minimizing the difference between the received observation and the reconstructed signal, the learned impairments converge to the true impairments. Although the computation of $\thetahat$ and $\tauhat$ in \eqref{eq:argmax_theta_tau} involves a nondifferentiable operation, it is possible to compute the gradient of the loss in \eqref{eq:loss_norm} with respect to $\bmkappahat$, which was already observed in an equivalent approach in \cite{yassine2022mpnet}.

\begin{algorithm}[tb]
    \caption{Unsupervised MB-ML of the gain-phase errors.}
    \label{alg:mb_learning}
    \begin{algorithmic}[1]
        \State \textbf{Input:} Initial \acp{GPI} $\bmkappa^{(0)}$.
        \State \textbf{Output:} Learned \acp{GPI} $\bmkappa^{(I)}$.
        \State \textbf{for} $i=1, 2, \ldots, I$
        \Indent
        \State Draw a batch of realizations of: $t, \gamma, \theta, \tau, \bmx$, and $\bmW$.
        \State Compute $\bmY$ according to \eqref{eq:Y_final}.
        \State Construct $\bmPhitheta(\bmkappa[i])$ and $\bmPhitau$ according to \eqref{eq:bmPhi_theta} and \eqref{eq:bmPhi_tau}.
        \State Compute the angle-delay map following \eqref{eq:ad_map}.
        \State Compute $\mathcal{J}(\bmkappa[i])$ according to \eqref{eq:loss_max} or \eqref{eq:loss_norm}.
        \State Update $\bmkappa^{(i+1)}$ through gradient descent.
        \State Normalize $\bmkappa^{(i+1)}$ so that $\norm{\bmkappa^{(i+1)}}^2 = N$.
        \EndIndent
    \end{algorithmic}
\end{algorithm}

In Algorithm~\ref{alg:mb_learning}, we summarize the proposed unsupervised MB-ML algorithm to learn the \acp{GPI}. The distributions of the random variables are highlighted in Table~\ref{tab:sim_parameters}. We initialize the algorithm with the ideal gain-phase coefficients, i.e., $\bmkappa^{(0)} = \bm{1}$.
Once we have learned the gain-phase errors according to Algorithm~\ref{alg:mb_learning}, we compute the same operations as the baseline in \eqref{eq:opt_test}, \eqref{eq:argmax_theta_tau} for inference, where the steering vector $\arx^\hermit(\theta;\bmkappa)$ is replaced by the steering vector with the learned impairments $\arx^\hermit(\theta;\bmkappa^{(I)})$.

\section{Results} \label{sec:results}
In this section, we detail the considered simulation parameters and present the sensing results\footnote{The code to reproduce all simulation results will be available in \url{github.com/josemateosramos/UL_gain_phase_ISAC} after the peer-review process.} to assess the effectiveness of the proposed learning approach.

\subsection{Simulation Parameters}
In Table~\ref{tab:sim_parameters} the simulation parameters are outlined, where we consider that the communication symbols $[\bmx]_s$ are randomly drawn from a \ac{QPSK} constellation and $\measuredangle(\cdot)$ denotes the phase of a complex value. The \ac{SNR} is $\SNR=\E[\norm{\gamma(\arx(\theta) (\bmb(\tau)\odot \bmx)^\top)}_F^2]/N_0 = \sigmagamma^2NS/N_0$. The magnitude and phase of the \acp{GPI} are drawn from the distributions detailed in \cite{jiang2013two}. To evaluate the objective function to maximize in \eqref{eq:opt_test} and \eqref{eq:argmax_theta_tau}, we perform a uniformly 2D grid search over angles and delays, similarly to \eqref{eq:ad_map}. During training, we leverage the Adam optimizer \cite{kingma2015adam}. 

\begin{table}[tb]
\setlength\extrarowheight{2pt}
\caption{Simulation parameters}
\label{tab:sim_parameters}
\begin{center}
\begin{tabular}{|p{.13\textwidth}|p{.13\textwidth}|p{.13\textwidth}|}
\hline
\textbf{Parameter} & \textbf{Expression}                  & \textbf{Value} \\ \hline
$N$  & - & 64 antenna elements \\ \hline
$f_c$ & - & 60 GHz \\ \hline
$\dR$ & $\lambda/2$ & 5 mm \\ \hline
$S$ & - & 256 subcarriers \\ \hline
$\Deltaf$ & - & 240 kHz \\ \hline
$\Tcp$ & $0.07/\Deltaf$ & - \\ \hline
$t$ & $\U\{0,1\}$ & - \\ \hline
$\gamma$  & $\cnormal(0,\sigmagamma^2)$ & -   \\ \hline
 $\theta$   & $\U[\thetamin, \thetamax]$                            &  -      \\ \hline
 $\thetamin$     & $\thetamean - \Deltatheta/2$ &  -     \\ \hline
 $\thetamax$ &  $\thetamean + \Deltatheta/2$&    -   \\ \hline
$\thetamean$ & $\U[-60\degree, 60\degree]$  & -      \\ \hline
 $\Deltatheta$ & $\U[10\degree, 20\degree]$  &  -     \\ \hline
 $\tau$ & $\U[\taumin, \taumax]$ & - \\ \hline
 $\taumin$ & $2\Rmin/c$  & $\Rmin=10$ m \\ \hline
 $\taumax$ & $2\Rmax/c$  & $\Rmax=43.75$ m \\ \hline
 $\bmx$ & $\E[\norm{\bmx}^2]=S$ & - \\ \hline
  $\vec(\bmW)$ &  $\cnormal(\bm{0}, N_0\bmI)$ &  - \\ \hline
  $\SNR$   & $\sigmagamma^2NS/N_0$     & 15 dB \\ \hline
  $|[\bmkappa]_n|$ & $\U[0.95,1.05]$ & - \\ \hline
  $\measuredangle([\bmkappa]_n)$ & $\U[-\pi/2, \pi/2]$ & - \\ \hline
  $\Ntheta, \Ntau$ & - & 100 \\ \hline
  Learning rate & - & $10^{-2}$ \\ \hline
  Batch size & - & $1024$ \\ \hline
  Training iterations & - & $10^4$ \\ \hline
\end{tabular}
\end{center}
\end{table}

\subsection{Impact of \acp{GPI}}
\begin{figure}[tb]
    \centering
    \includegraphics[width=0.5\textwidth]{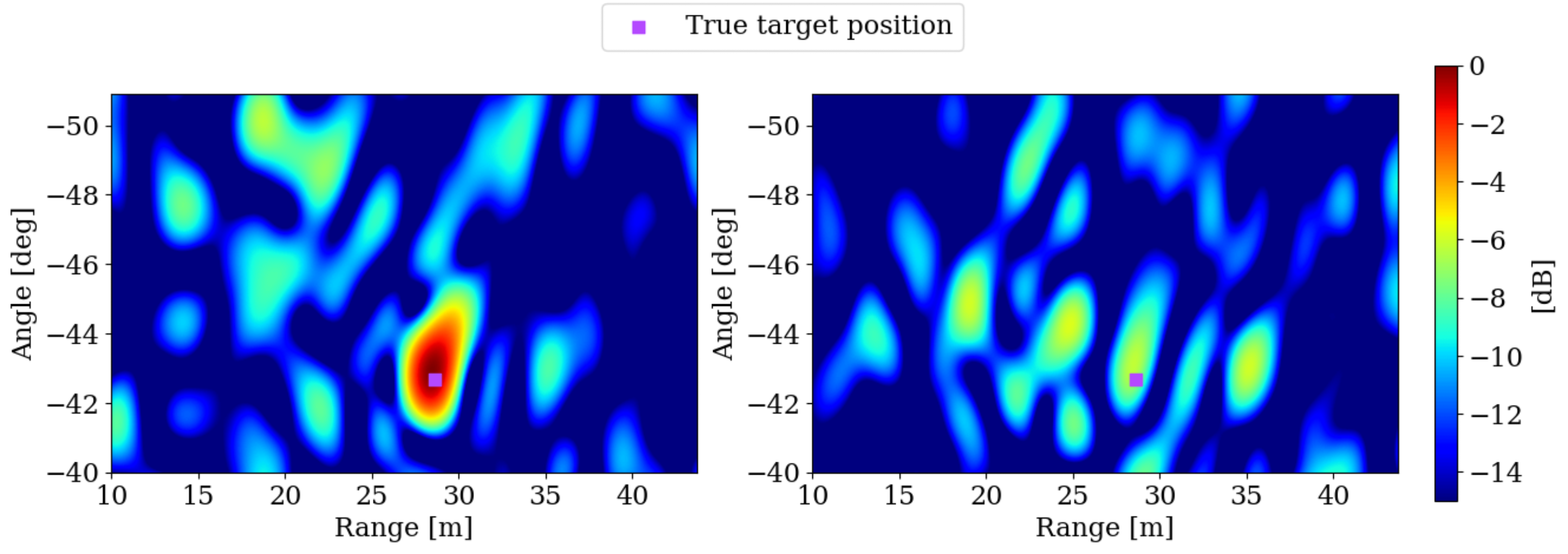}
    \caption{Angle-delay map under full knowledge of the impairments (left) and assuming no impairments (right). The images are normalized with respect to the maximum of the angle-delay map under full knowledge of the impairments.}
    \label{fig:ad_map}
\end{figure}

To understand how ignoring the \acp{GPI} affects the sensing performance, we plot in Fig.~\ref{fig:ad_map} the angle-delay maps under full knowledge of the \acp{GPI}, i.e., $\bmkappahat=\bmkappa$ (left) and assuming no impairments, i.e., $\bmkappahat = \bm{1}$ (right). The channel model in \eqref{eq:Y_final} includes an impairment realization $\bmkappa \neq \bm{1}$. It is observed that disregarding \acp{GPI} changes the position of the maximum of the angle-delay map, which affects the test in \eqref{eq:argmax_theta_tau}.
Furthermore, the angle-delay maps in Fig.~\ref{fig:ad_map} are normalized with respect to the maximum of the angle-delay map under full knowledge of \acp{GPI} (left), which implies that ignoring the \acp{GPI} also decreases the maximum magnitude of the angle-delay map compared to full knowledge of \acp{GPI} (as commented in Sec.~\ref{sec:proposed_method}).

\subsection{Sensing Results} \label{subsec:sensing_results}

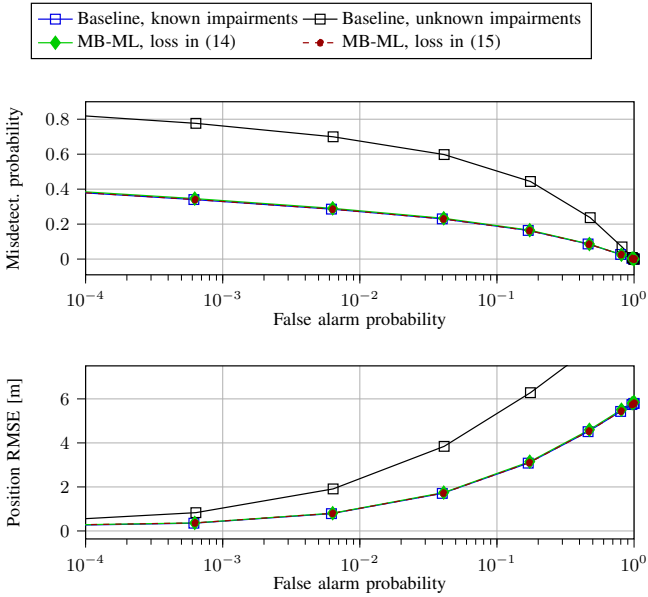
\begin{figure}[tb]
    \centering
    \begin{tikzpicture}[font=\small]

\begin{axis}[
width=9.5cm,
height=3cm,
scale only axis,
legend cell align={left},
legend columns = 2,
legend style={
  fill opacity=1,
  draw opacity=1,
  text opacity=1,
  at={($(current axis)+(0.4,3cm)$)}, 
  anchor=west
},
log basis x={10},
tick align=outside,
tick pos=left,
x grid style={white!69.0196078431373!black},
xlabel={False alarm probability},
xmajorgrids,
xmin=1e-4, xmax=1,
xtick style={color=black},
y grid style={white!69.0196078431373!black},
ylabel={Misdetect. probability},
ymajorgrids,
ymax=0.9,
xmode=log,
ytick style={color=black},
]

\addplot[color=blue!90!black, mark=square, mark options={scale=1.2}, name path=A] table {figures/data_roc/baseline_ideal_roc.txt};
\addlegendentry{Baseline, known impairments};

\addplot[color=black, mark=square, mark options={scale=1.2}, name path=B] table {figures/data_roc/baseline_gp_roc.txt};
\addlegendentry{Baseline, unknown impairments};

\addplot[color=green!80!black, mark=diamond*, mark options={scale=1.6}] table {figures/data_roc/learning_max_map_roc.txt};
\addlegendentry{MB-ML, loss in (14)};

\addplot[color=red!60!black, mark=*, dashed, mark options={scale=0.8}] table {figures/data_roc/learning_reconst_roc.txt};
\addlegendentry{MB-ML, loss in (15)};

\end{axis}

\begin{axis}[
width=9.5cm,
scale only axis,
height=3cm,
at={(0,-1200)},     
log basis x={10},
tick align=outside,
tick pos=left,
legend columns = 1,
legend style={
  fill opacity=1,
  draw opacity=1,
  text opacity=1,
  at={(0.5,1.45)},
  anchor=center
},
x grid style={white!69.0196078431373!black},
xlabel={False alarm probability},
xmajorgrids,
xmin=1e-4, xmax=1,
xtick style={color=black},
y grid style={white!69.0196078431373!black},
ylabel={Position RMSE [m]},
ymajorgrids,
ymax=7.5,
xmode=log,
ytick style={color=black}
]

\addplot[color=blue!90!black, mark=square, mark options={scale=1.2}] table {figures/data_roc/baseline_ideal_pfa_pos_rmse.txt};

\addplot[color=black, mark=square, mark options={scale=1.2}] table {figures/data_roc/baseline_gp_pfa_pos_rmse.txt};

\addplot[color=green!80!black, mark=diamond*, mark options={scale=1.6}] table {figures/data_roc/learning_max_map_pfa_pos_rmse.txt};

\addplot[color=red!60!black, mark=*, dashed, mark options={scale=0.8}] table {figures/data_roc/learning_reconst_pfa_pos_rmse.txt};

\end{axis}

\end{tikzpicture}
    \caption{Sensing results as a function of the false alarm probability.}
    \label{fig:sens_results}
\end{figure}

In Fig.~\ref{fig:sens_results}, we show the testing sensing results, where we compare: (i) the baseline of Sec.~\ref{sec:baseline} when $\bmkappa$ matches the true impairments (in blue), (ii) the baseline with $\bmkappa = \bm{1}$ (in black), and (iii) the proposed method of Sec.~\ref{sec:proposed_method} with the learned impairments $\bmkappa = \bmkappa^{(I)}$, using the loss in \eqref{eq:loss_max} (in green) and the loss in \eqref{eq:loss_norm} (in red). The results in Fig~\ref{fig:sens_results} are averaged over 100 realizations of the \acp{GPI}. 
The results in Fig.~\ref{fig:sens_results} indicate that the proposed unsupervised learning approach can converge to a \ac{GPI} vector similar to the true impairments of the ULA and obtain similar performance as the case where the impairments are fully known. This confirms the hypothesis of Sec.~\ref{sec:proposed_method} about the effectiveness of the proposed loss functions.
Moreover, the performance of the algorithm using the losses \eqref{eq:loss_max} and \eqref{eq:loss_norm} is very similar, which can be explained following the derivation of Appendix~\ref{app:maprt}. The loss in \eqref{eq:loss_norm} resembles the objective to minimize in \eqref{eq:maprt_3} and the loss in \eqref{eq:loss_max} is similar to \eqref{eq:maprt_final}. Both \eqref{eq:maprt_3} and \eqref{eq:maprt_final} are derived from the same MAPRT objective in \eqref{eq:maprt_1}. The results in Fig.~\ref{fig:sens_results} indicate that minimizing \eqref{eq:loss_max} or \eqref{eq:loss_norm} is equivalent. The advantage of \eqref{eq:loss_max} is that is does not require knowledge of the SNR, while \eqref{eq:loss_norm} can be more easily generalized to multiple targets and embedded in iterative algorithms like the orthogonal matching pursuit algorithm \cite{OMP_mmWave_2016}.

\section{Conclusions} \label{sec:conclusions}
In this work, we have proposed a model-based unsupervised learning approach to account for \acp{GPI} in the RX ULA of an ISAC system. We have based our proposed approach on the optimal MAPRT, developing a differentiable approach that allows for backpropagation, and proposing two unsupervised loss functions that require no labeled data. Our results have shown that the proposed approach can effectively compensate for the effect of gain-phase errors in the RX ULA, yielding target detection and position estimation performances similar to the case where the impairments are fully known. Natural extensions include considering multiple targets and the \ac{GPI} under a MIMO system. 

\appendices

\section{Derivation of MAPRT} \label{app:maprt}
This appendix details the derivation of the MAPRT in Sec.~\ref{sec:baseline} for single-target detection and position estimation.
Since the target in the far-field is randomly present, we can formulate the target detection problem as a binary hypothesis testing problem:
\begin{align}
\begin{aligned}
    & \mathcal{H}_0: \bmY = \bmN \\
    & \mathcal{H}_1: \bmY = \gamma\bmM(\theta, \tau) + \bmN,
\end{aligned}
\end{align}
where $\bmM(\theta, \tau) = \arx(\theta;\bmkappa) (\bmb(\tau) \odot \bmx)^\top$. 

Note that the transmitted communication symbols $\bmx$ are known for the sensing receiver in the considered monostatic setup. Considering $\gamma$ a random unknown, the MAPRT is
\begin{align} \label{eq:maprt_1}
    \Lcal(\bmY) = \frac{\max_{\gamma, \theta, \tau}p(\gamma, \theta, \tau, \Halt|\bmY)}{p(\Hnull|\bmY)} \hdet \eta.
\end{align}
Applying the Bayes' theorem to \eqref{eq:maprt_1} yields
\begin{align} \label{eq:maprt_2}
    \Lcal(\bmY) = \frac{\max_{\gamma, \theta, \tau}p(\bmY|\gamma, \theta, \tau, \Halt) p(\gamma)p(\theta)p(\tau)p(\Halt)}{p(\bmY|\Hnull)p(\Hnull)} \hdet \etatilde.
\end{align}
Assuming $p(\Hnull)=p(\Halt)=1/2$, $\gamma\sim\cnormal(0,\sigmagamma^2), \theta\sim\U([\thetamin, \thetamax]), \tau\sim\U([\taumin, \taumax])$ and taking the logarithm in \eqref{eq:maprt_2}, we obtain\footnote{We consider $\bmM \triangleq \bmM(\theta,\tau)$ for convenience.}
\begin{align} \label{eq:maprt_3}
    \Llog(\bmY) = \frac{\norm{\bmY}_F^2}{N_0} - \min_{\substack{\gamma \\ \theta\in\Oset \\ \tau\in\Tset}} \bigg\{ \frac{\norm{\bmY-\gamma\bmM}_F^2}{N_0} + \frac{|\gamma|^2}{\sigmagamma^2}\bigg\} \hdet \etabar,
\end{align}
where $\Llog(\bmY) = \log(\Lcal(\bmY))$ and $\etabar=\etatilde + \log(\pi\sigmagamma^2) + \log(\thetamax-\thetamin) + \log(\taumax-\taumin)$. The optimal $\gamma$ for a given $(\theta, \tau)$ in \eqref{eq:maprt_3} is
\begin{align} \label{eq:gamma_hat_1}
    \gammahat = \frac{\vec(\bmM)^\hermit\vec(\bmY)}{\norm{\bmM}_F^2 + N_0/\sigmagamma^2}.
\end{align}
Manipulating the expression in \eqref{eq:maprt_3} and plugging \eqref{eq:gamma_hat_1} yields
\begin{align} \label{eq:Llog_final}
    \Llog(\bmY) = \max_{\substack{\theta\in\Oset \\ \tau\in\Tset}} \bigg\{\frac{|\vec(\bmM)^\hermit\vec(\bmY)|^2}{\norm{\bmM}_F^2 + N_0/\sigmagamma^2} \bigg\} \hdet \etabar.
\end{align}
Given the definition of $\bmM(\theta, \tau)$, we have that
\begin{align}
    \norm{\bmM}_F^2 &= N \norm{\bmx}_2^2 \label{eq:norm_M_ideal} \\
    |\vec(\bmM)^\hermit \vec(\bmY)|^2 &=  |\arx^\hermit(\theta;\bmkappa) \bmY (\bmb(\tau) \odot \bmx)^\ast|^2. \label{eq:vec_M_vec_Y_ideal}
\end{align}
Plugging \eqref{eq:norm_M_ideal} and \eqref{eq:vec_M_vec_Y_ideal} into \eqref{eq:Llog_final} yields
\begin{align} \label{eq:maprt_final}
    \Llog(\bmY) = \max_{\substack{\theta\in\Oset \\ \tau\in\Tset}} \big\{ |\arx^\hermit(\theta;\bmkappa) \bmY (\bmb(\tau) \odot \bmx)^\ast|^2 \big\} \hdet \eta,
\end{align}
where $\eta = \etabar (N \norm{\bmx}_2^2 + N_0/\sigmagamma^2)$. Once we have obtained the estimated $\thetahat, \tauhat$, we can plug the estimations in \eqref{eq:gamma_hat_1}.
\balance 

\bibliographystyle{IEEEtran}
\bibliography{references}

\end{document}